# Effects of Nanodots Shape and Lattice Constants on the Spin Wave Dynamics of Patterned Permalloy Dots

Nikita Porwal[1], Jaivardhan Sinha[2] and Prasanta Kumar Datta[1]*

[1]Department of Physics, Indian Institute of Technology Kharagpur, W.B. 721302, India
[2]Department of Physics and Nanotechnology, SRM Institute of Science and Technology, Kattankulathur 603203, Tamil Nadu, India
*Email: *pkdatta@phy.iitkgp.ac.in

Micromagnetic simulations studies on Permalloy ($Ni_{80}Fe_{20}$) nanodot with different shape and edge-to-edge separation (*s*) down to 25nm arranged in square lattice are reported. We observe the significant variation of spin-wave (SW) dynamics of nanodots of different shapes (triangular, diamond and hexagon) and of fixed dot diameter 100nm with varying *s*. Modes for single dot are transformed in an array into multiple quantized, edge and centre modes for different shapes and edge-to-edge separations, with different spin wave frequencies and peak intensities. Specifically, in the triangular dot sample, a broad range of mode frequencies is observed with highest SW frequency 14.7 GHz. For separation less than 100nm, the SW frequencies undergoes significant modification due to the varying nature of the magnetostatic and dipolar interaction in the array while for separation above 100nm, the SW frequency mostly remain constant. The power profiles confirm the nature of the observed modes. The spatial profiles of magnetostatic field are determined by a combination of internal magnetic-field profiles within the nanodots and the magnetostatic fields within the lattice. The inter-dots interaction of magnetostatic field shows dipolar and quadrupole contributions for all the shapes. Interestingly, vortex states with shifted core and polarity are observed in the array for all the shapes at $H_{bias}$ = 0. Our results provide important understanding about the tunability of SW spectra in the array of triangular, diamond and hexagon shaped nanoelements.

**Keywords:** Magnonic crystal, Spin wave, Micromagnetic simulation

**Introduction**

There has been a growing interest in exploring the magnetization properties and spin wave (SW) dynamics of magnetic nano-elements, due to its potential applications in magnetic logic devices [1, 2, 3, 4], magnetic resonance imaging [5], as well as in magnetic storage [6]. The field has gained a boost due to advancements in nanofabrication and characterization techniques [7]. Recently, magnetic bit patterned media (BPM) based on nanostructures are found to be promising candidates over magnetic thin films since there is no temporal overlapping of information between the bits and thus allowing non-interactive storage of bits [8]. In patterned nanostructures, the individual nano-elements magnetostatically interact with the neighbouring ones forming characteristic spin waves with mode frequencies dependent on shape and size of the structures [9]. In magnonic crystals, nanomagnets encounter magnetic coupling through magnetostatic stray fields and thus give rise to different SW modes due to collective behaviour of the SW propagation throughout the whole array [10, 11, 12, 13, 14]. For large values of lattice constants, the collective behaviour of the dots gets terminated and thus results in individual dynamics of the nanodots. To explore the potentiality of these multitasking nanodots in novel computing devices, studies of their static and dynamic magnetization properties have become important [15, 16, 17].

The most attractive and valuable features of nanomagnets are their different ground states under certain conditions. The onion, buckle, flower and vortex states are main configurations in magnetic nanodots. The demagnetization field plays a key role in favouring the vortex configuration in low external fields. A variety of analytical and computational methods such as energy functional theory, variational theory and micromagnetic simulations, are reported to clarify the quasi-single domain, vortex states and spin wave in thin film [18, 19]. Experiments on individual nanomagnets below 200nm have been reported in refs. [13, 14, 20] which focus on either the vortex state or the quasi single domain state. Very few reports are found on the time-resolved spin-wave dynamics of nanodots of diameter less than 100nm [21, 22]. A dissipative soliton dynamics has been studied in a nano-dot chain of diameter 90nm [23]. It has been reported that the simplest system for studying static and dynamic magnetic properties of nanomagnets is the nano-ellipsoid [24] because of its shape and can be characterized by a uniform magnetization and a homogeneous internal field. Many reports are there on the planar nanostructures with a variety of shapes (e.g. rectangles, square, circles, ellipses, rings, anti-rings, triangles) for studying their static and dynamic properties characterized by anisotropy [9, 25, 26]. The complicated effects of the anisotropic nanomagnets on the spin-wave spectra were experimentally determined by Brillouin light scattering (BLS) measurements in circular rings [27], where splitting and localization of spin excitations were found to be related to the in-homogeneity of the internal field. In the case of triangular dot, the problem is even more complicated and only few attempts were made for studying static and dynamics properties [28, 29] and the vortex core dynamics [30]. Triangular rings in onion and vortex states have been explored in experimental and simulated hysteresis loops by diffracted magneto-optical Kerr effect (D-MOKE) [31, 32]. The vortex state associated with the dot triangular shapes was examined for high density memories because of very low stray fields. A systemic study of different ground states in equilateral triangles was first reported in ref. [33, 34]. Depending upon the direction of the magnetizing field, the magnetization at saturation exhibits a Y- state when the external field is applied along the height of the triangle or a "buckle-state" when the external field is applied along the base. At zero bias field, depending upon the aspect ratio, the vortex state can be another possibility. The spin-wave modes of micron-sized triangular dots have also been studied in ref. [31, 32, 33]. In general, the experimental and theoretical studies on magnetic nanodots of circular and square shape with separation down to 50nm has been performed in past and reasonable good agreement between theory and experiments have been obtained. Due to limitation of lithography process, mostly the corners of the nanodots with triangular shape get rounded and it becomes difficult to exactly decipher the role of sharp corners on the spin-wave spectra. Based on the above review, it is evident that there is hardly any comprehensive report on the hexagonal, diamond and triangular patterned magnetic material with different inter-dot separations for spin wave dynamics. It is worth mentioning here that the detailed SW mode characteristics of the hexagonal shape individual nanodots are missing to best of our knowledge.

In this work, we present a systematic micromagnetic simulation study of spin-wave dynamics of Py dot of, triangular diamond, and hexagonal shapes. We investigate the spin wave dynamics of the samples from single dot to array of square lattice with varying $s$ from 25nm to 500nm. The dot diameter ($d$) taken in this study is as low as 100nm, which has not been studied before for the shapes considered here [9, 12, 35, 36]. A simple square lattice is chosen for all shapes of nano-dot arrays to avoid exotic inter-dot interaction mediated by the dynamic stray magnetic field observed for different lattice symmetries. We find that the single-dot samples show few SW modes while dot-array samples show multiple modes (edge, center, quantized and extended). The precessional frequencies show a significant variation with varying shapes and edge-to-edge separations due to the change in magnetostatic field in the array. We observe a clear variation in the SW dynamics associated with a gradual transition from a strongly collective to a completely isolated dynamical behaviour with the variation of $s$. Also, the variation of magnetostatic field with $s$ is understood by considering both dipolar and quadrupole interactions. We find different magnetic configurations such as (i) onion state with bias field $H_{bias}$ = 1.12kOe in, diamond and hexagonal dots, (ii) buckle state with $H_{bias}$ = 1.12kOe in triangular dots and (iii) vortex state with $H_{bias}$ = 0 in all the shapes. Our work can provide a guideline for the fabrication of nanomagnets for efficient magnonic devices.

**Method**

We study arrays of different shapes (triangular diamond and hexagonal) of Py nano-dots with 100nm diameter and varying edge-to-edge separation from 25nm to 500nm. The arrays are made up of 5 × 5 elements, time dependent micromagnetic simulations are performed by solving the Landau-Lifshitz-Gilbert (LLG) equation using the public domain software object oriented micromagnetic framework (OOMMF) [37] and LLG micromagnetic simulator (V4, Mindspring, M. R. Scheinfein, year of release 2015) [38]. Calculations are carried out by dividing the samples into a two-dimensional array of cells with dimensions of 2.5nm × 2.5nm × $t$ nm, considering two-dimensional periodic boundary conditions (2-D PBC) [39] for all the samples, where $t$ = 20nm is the thickness of the sample. Details on doing micromagnetic simulations with OOMMF can be found in ref. [40]. The linear dimension of the cells is less than the exchange length, which is defined as $\sqrt{(A/2\pi M^2)}$, where *A* is the exchange constant and *M* is the magnetization respectively, and its value is 5.3nm for Py.

The static magnetic results are obtained by first applying a large enough bias field to fully magnetize the sample and then reducing the magnetic field to the bias field value. During this process, the damping coefficient, α is set to 0.9 so that the precession dies down quickly and the magnetization fully relaxes within 4ns. By this way, the system is allowed to reach the equilibrium. The magnetization dynamics is simulated by applying a perpendicular pulsed magnetic field *h(t)* with rise time of 10ps and peak value of 30Oe, as shown in fig. 1(a), and α is set to 0.008. The simulations assume typical material parameters for Py, gyromagnetic ratio *γ* = 18.5MHz/Oe, magneto-crystalline anisotropy $H_k = 0$, saturation magnetization $M_s$ = 860 emu/cc, and exchange stiffness constant *A* = 1.3×10$^{-6}$ erg/cm [41]. Calculations for the spatial maps of magnetization at time steps of 10ps are performed for total time duration of 4ns. We mimic the configuration of time-resolved MOKE in the simulations. The time variation of magnetization of each cell is recorded. The Fourier transform of the time variation gives the spin wave spectra. The Fourier transform images, consisting of amplitude and phase profile of the magnetization of the complete sample, are obtained for each mode frequency by putting together the Fourier transform spectra of the cells.

**Results and Discussions**

We investigate the dependence of the static and time-resolved magnetization dynamics on the Py dot samples with varying shapes (triangular, diamond and hexagonal) and varying *s* from 25nm to 500nm with fixed diameter *d* = 100nm at fixed thickness *t* = 20nm as shown in fig. 1(a). Figure 1(b) shows the static magnetic configuration of the dot array samples for *s* = 100nm. Figure 1(c) and (d) show the time-resolved magnetization ($M_z$) and corresponding fast Fourier transform (FFT) of a single diamond dot sample of *d* = 100nm respectively. In-plane bias field of $H_{bias}$ = 1.12kOe is applied in the geometry as shown in fig. 1(a). The strength of the bias field is taken much above the saturation field of Py. We observe that the edges parallel to the bias field show yellow and pink regions at the dots edges due to demagnetization of magnetic field as shown in fig. 1(b). Figures 2(a) and (b) show the time-dependent magnetization precession ($M_z$) and the corresponding FFT spectra for single dot samples with dot diameter 100nm respectively. The static magnetic configurations of samples are shown in the inset of fig. 2(b). We observe that magnetization precession vary with the dot shapes and thus give different SW modes. A bias field of $H_{bias}$ = 1.12kOe is applied to ensure saturated static magnetization for the single elements. The static magnetic configurations will be discussed in detail later. The time-resolved magnetization and the corresponding FFT spectra show SW precession modes of the single elements. In the case of triangular dot, we observe three modes with dominant one at 3.2GHz (mode 3) and higher frequency modes (mode 2 and 1) at 9.4 and 11.9GHz. The demagnetized regions (red and blue regions) near the edges of the dots indicate the non-uniform fields, which are responsible for forming the observed modes. The diamond dot samples show two frequency modes, a sharp peak at 5.7GHz (mode 2) and higher frequency mode with lower intensity at 13.2GHz (mode 1). However, the hexagon dot sample shows single uniform precession mode at 5.7GHz.

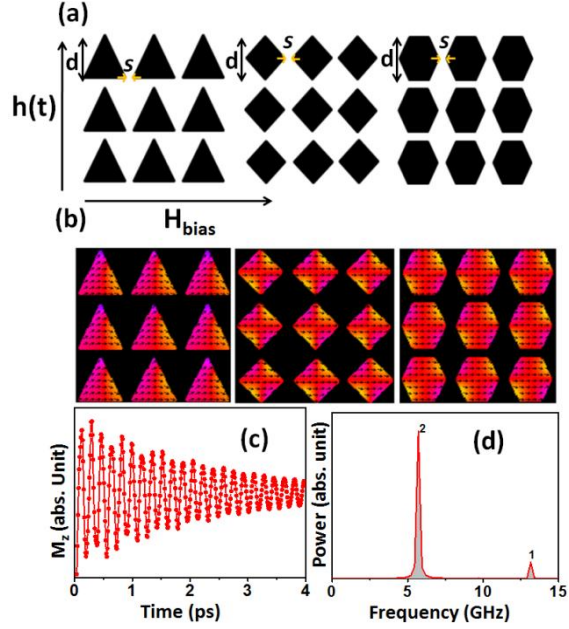

Figure 1 (a) Geometry of simulation of time resolved dynamics for Py dot array of different shapes with dot diameter $d = 100$nm, same thickness $t = 20$nm, arranged in a square lattice with varying edge-to-edge separation $s$ from 25nm to 500nm, (b) simulated static magnetic configurations of the dot array samples for $s = 25$nm, (c) time dependent magnetization and (d) corresponding FFT spectra are shown for a single diamond dot with 20nm thickness. A bias field $H_{bias}= 1.12$ kOe is applied in the geometry as shown in (a).

In order to understand how the SW modes are modified in an array with varying $s$, we plot the power spectra of the samples for $s = 25$, $s = 100$, $s = 300$ and $s =500$nm in fig. 3(a), (b), (c) and (d) respectively. There are remarkable differences in the SW spectra in the arrays from that of the single nano-dot. The triangular dot array sample shows nine modes, within a broad range of frequencies starting from 1.3 to 14.7GHz, unlike only three modes for a single triangular dot. As we have seen that in an array, the width of the peaks is reduced from that of the single dot. Also, the peak intensities and frequencies are varied significantly in array samples. The diamond-shaped dot array shows five modes. For hexagonal shape, both the single dot and the array of dots exhibit only one mode, at 5.7GHz for the single dot and at 6.9GHz for the array. When $s$ is varied from 25nm to 100nm, we observe variations in frequency modes for samples of any shape. Specifically, there is a sudden drop in frequency of modes due to the change in the collective nature of the SW dynamics. The diamond dot array exhibits six distinct frequency modes for $s = 100$nm and five modes for $s = 25$nm. In hexagonal dot array, a single mode (6.9GHz) for s = 25nm is split into two modes, mode 1 and 2 (6.7 and 6.0GHz) at $s = 100$nm respectively.

The triangular dot array ($s = 100$nm) shows ten modes covering a broad range of frequency spectrum from 1GHz to 13.6GHz similar to the nine modes observed for $s = 25$nm. The observation of large number of modes for triangular shape and diamond shape is qualitatively consistent to the experimental and theoretical findings in ref. [9]. Slight variation in mode frequencies in comparison to the present work is likely due to the difference in the size of the dot some rounding off the corners of triangular and diamond.

From figure 3(b), (c) and (d) we observe that for diamond dot array, modes 1 and 2 (11.6 and 11.2GHz) at $s = 100$nm maintain their frequency at $s = 300$nm while mode 3 (6.7GHz) of $s = 100$nm, slightly shifted to lower frequency side (6.5GHz) at $s = 300$nm. For $s = 500$nm, we observe mode 2 (6.2GHz) is the highest intensity mode similar to the mode 3 of $s$ = 300nm with frequency shifted to lower side. In the case of hexagonal dot array, we observe mode 1 and 2 maintain their frequency for $s$ = 300nm and $s$ = 500nm separations. In triangular dot array, all the modes maintain their frequencies for all the samples. Apart from this, the mode frequencies remain almost same for $s$ = 300 to 500nm while the relative intensities and line widths of the modes changes with the increase in inter-dot separation. We observe that the mode numbers are getting decreased with increasing $s$, except triangular shape which shows ten modes for $s$ = 500nm and nine modes for $s$ = 300nm. This is due to the decrease in the areal density of the nano-elements in the array.

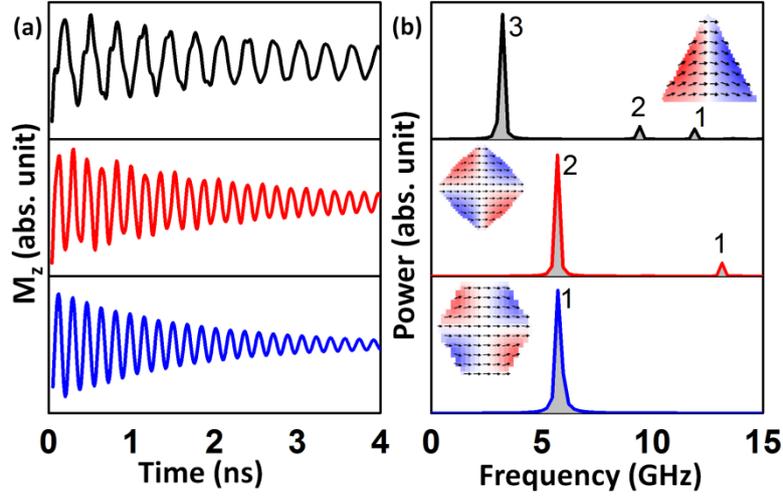

Figure 2. (a) Time-dependent magnetization and (b) corresponding FFT spectra for single Py dot samples with fixed dot diameter 100nm and thickness 20nm for different geometries (inset of fig. (b)).

In fig. 4, precession frequencies of arrays of Py nanodots are plotted as a function of $s$ for different shapes (a) triangular, (b) diamond, and (c) hexagonal samples. We observe that precession frequency modes of arrays increase with the decrease in $s$. Similar, trend for triangular and diamond dots has also been observed in earlier experimental report [9]. This may be due to strong dipole interaction between the dots at lower edge-to-edge separations. However, the proximity of the elements increases the magnetostatic field interaction which overwhelms the demagnetizing effect. Thus, the effective magnetic field increases due to the suppression of demagnetization field resulting in observed increase of precession frequency with decreasing $s$. Moreover, there are some exceptions such as for diamond and hexagon dot array, which show higher frequency modes for larger $s$ values. This may be attributed to the shape of the dots. For $s$ = 25nm, a number of precession modes appear as we have seen in fig. 3(a). These are the collective modes of precession in an array [12] whose explanation will be described in detail using the magnetostatic stray field contour plots later in this paper.

To analyze the different SW modes, we did spatially and frequency resolved FFT imaging of the SW modes using LLG micromagnetic simulator [38]. The power distributions of the modes for the array nanodots samples having $s$ = 100nm are shown here in fig. 5. As we observe from fig. 4, that all the samples with edge-to-edge separation 100nm to 500nm, show almost same mode frequency and thus show the same nature as that of nanodots at $s$ = 100nm. The triangular dot array shows different mode profiles as opposed to the other shapes. Here, mode 1, 2 and 3 shows BV like standing SW mode with quantization number n = 9, 7 and 5 respectively. Mode 4 shows edge mode like behaviour. While, all other lower frequency modes 5-10 show edge modes with non-uniform distribution of the power at the edge of the dots at alternate column of the array [9, 15]. For the diamond dot array, mode 1 and mode 2 show quantized SW mode with mixed quantization numbers n = 3 and n = 5 in alternate columns. Mode 3 corresponds to centre mode and mode 4 shows well known edge mode of the array where power is confined at the edges of the dots. Mode 5 and 6 are not uniform throughout the whole array and show centre mode like behaviour. For the hexagonal dot arrays, mode 1 shows quantized SW mode with quantization number n = 3 and n = 5 respectively while, mode 2 corresponds to edge mode. To gain more insight into the dynamics, we have simulated the magnetostatic field distribution of the arrays by using LLG micromagnetic simulator. The contour plots of 3 × 3 elements from the centre of the large array are shown in fig. 6 with varying $s$ (a) 25nm, (b) 100nm, (c) 300nm and (d) 500nm. We observe that density of the interacting field lines reduces remarkably with the increase in $s$ and nearly vanishes for higher $s$ for all shaped samples. To evaluate the inter-dot interaction, we took line scans along the dotted white lines as shown in Fig. 6 (a)-(d). For $s$ = 25nm, the magnitude of stray field between two consecutive dots is maximum for all shapes samples, which sharply decreases with the increase in the $s$ and becomes negligibly small for $s$ = 500nm separation. We observe that at $s$ = 25nm, nanodots are strongly coupled by magnetostatic field interactions and thus giving rise to multiple collective modes of the array. Figure 6(e) shows the variation of magnetostatic field with the lattice constants for all the samples. To quantify the inter-dot interaction, we took line scans along the dotted white lines for all edge-to-edge separations and the data fits with consideration of both dipolar and quadrupole interactions terms. We observed that the inter-dots interaction is a mixture of dipolar and quadrupole in nature for all the shapes as opposed to the quadrupole interaction dominated than the dipolar seen in the early report for square dot array [12]. From here, it is clear that at *s*= 25nm, the system is strongly coupled via magnetostatic interaction which can guide a broad spectrum of long wavelength SW to carry and progress information. Hence, they can be considered as a promising candidate for magnonics applications. Further to understand the origin of the change in spin wave modes with changing dot shapes we have plotted the magnetostatic field distribution with distance of the sample for all nanodots arrays. Figure 7(a), (b) and (c) shows the contour maps of triangular, diamond and hexagon samples. We took the line scan (from dotted white line region) of the sample. The magnetostatic field shows maxima and minima across the two different regions A (green dotted box) and B (brown dotted box) for different shaped samples (fig. 7 (d)). One can readily observe that the magnetostatic field strength is increases with changing shapes at region B and at region A. It shows different strength, which is due to the different shapes of the dots, as different shapes experience different magnetostatic field profile, and the observed magnetostatic field value is shown in table 1. We observe less magnetostatic field may be due to the small size of the nanodots of 100 nm than that of 200 nm and 250 nm reported earlier in ref. [9, 15, 35].

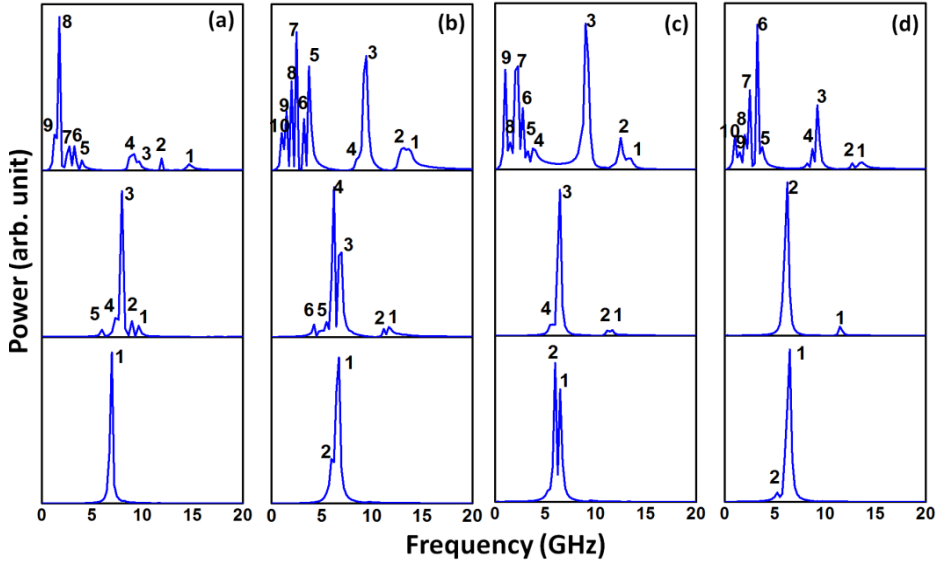

Figure 3. Spin wave spectra for arrays of Py nanodots with (a) $s$ = 25nm, (b) $s$ = 100nm, (c) $s$ = 300 nm and (d) $s$ = 500nm are shown. Different rows in the figure are for triangular, diamond and hexagonal shapes (top to bottom).

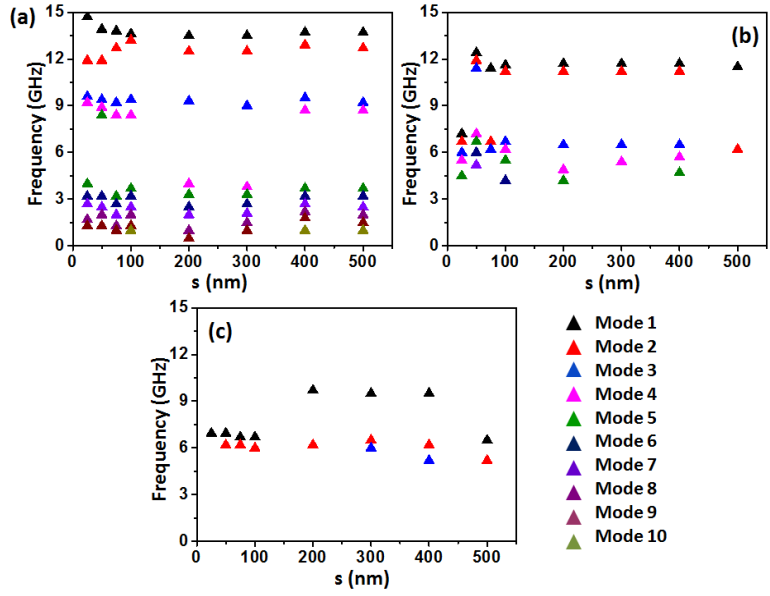

Figure 4. The precession frequency of arrays of Py nanodots are plotted as a function of *s* for different shapes (a) triangular (b) diamond and (c) hexagonal samples.

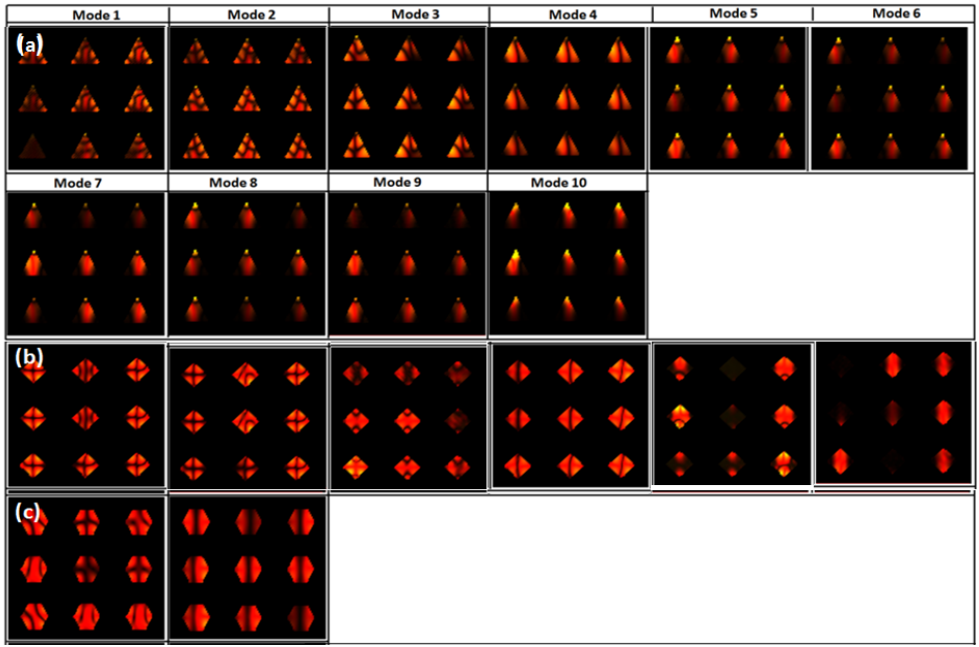

Figure 5. Simulated spin wave mode profiles from the central part of the arrays of 7 x 7 of nanodots with lattice constant 100nm of (a) triangular, (b) diamond and (c) hexagon shaped nanodots samples. The mode numbers are given at the top of the mode profiles.

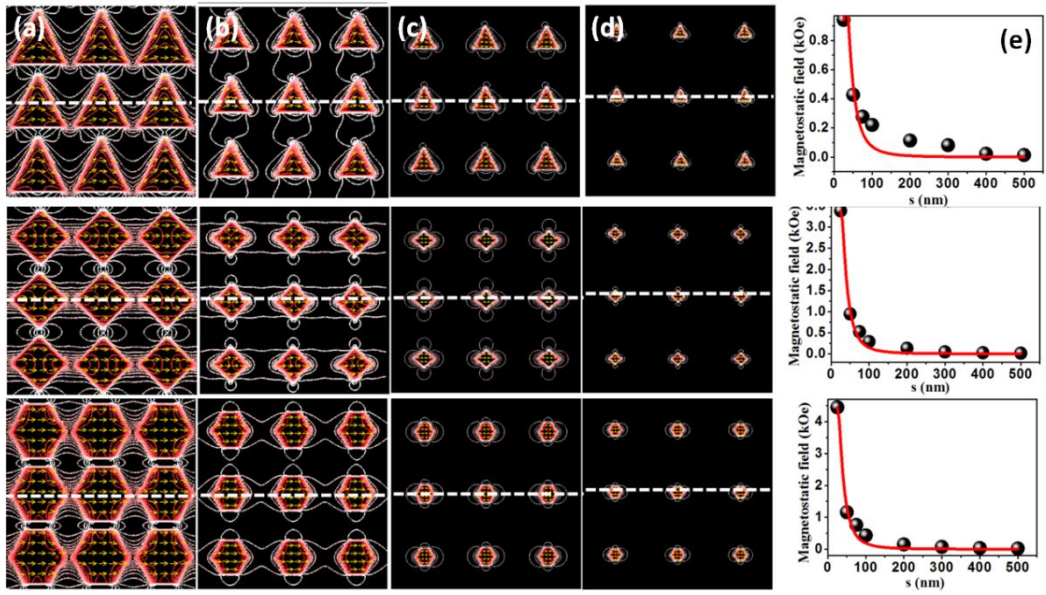

Figure 6. The contour maps of simulated magnetostatic field distribution from the central part of the arrays of 7×7 dots with triangular, diamond and hexagonal shape with varying edge-to-edge separation of (a) 25nm, (b) 100nm, (c) 300nm and (d) 500nm respectively are shown. (e) The variation of magnetostatic stray field with the lattice constant (circular symbols: micromagnetic simulation, solid line: fitted curve. The bias field of 1.12kOe was applied along the x- axis. The arrows inside the dots represent the magnetization states of the dots and the strength of magnetostatic field is represented by the colour bar given at the right side of the figure. The white horizontal lines represent the position of the lattice from where the line scans have been taken.

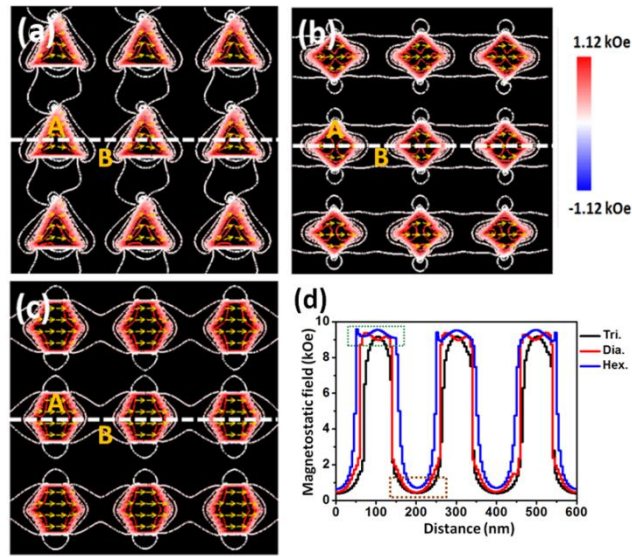

Figure 7. The contour maps of simulated magnetostatic field distribution (x-component) from the central part of the arrays of 7×7 dots of (a) triangular, (b) diamond and (c) hexagonal shape for s = 100nm are shown. The bias field of 1.12kOe was applied along the x- axis. The arrows inside the dots represent the magnetization states of the dots and the strength of magnetostatic field is represented by the colour bar given at the left side of the figure. The white horizontal lines represent the position of the lattice from where the line scans have been taken. Comparison of the simulated magnetostatic field distribution in different shapes for *s* = 100nm is shown in (f) taken along the dotted lines from samples.

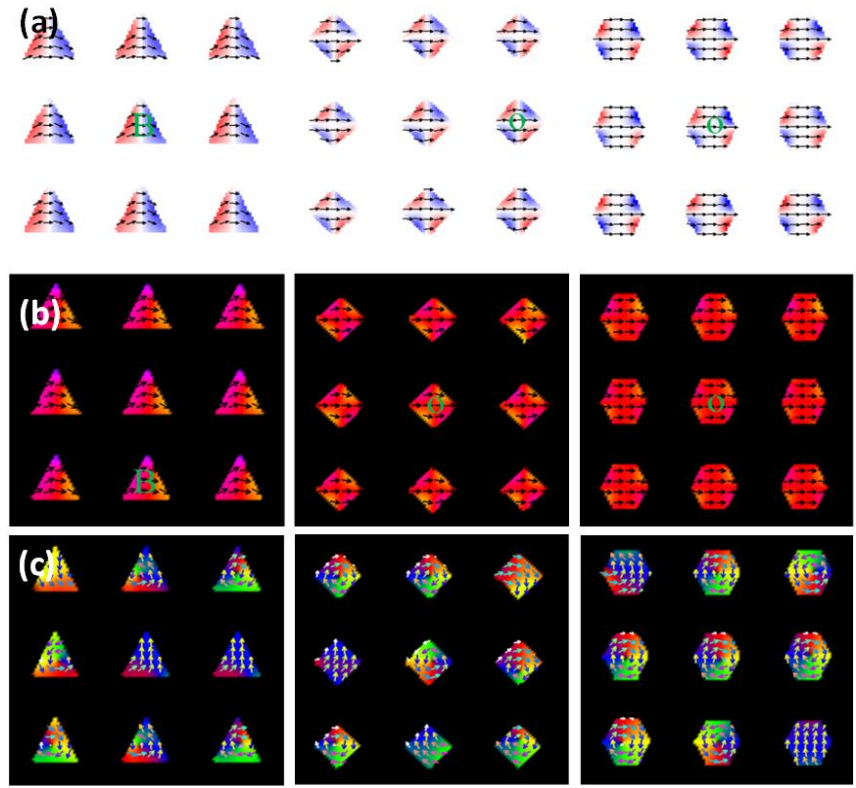

Figure 8. The ground state magnetization images from arrays of Py nanodots samples with s = 100nm and thickness 20 nm are simulated using (a) OOMMF, (b) LLG micromagnetic simulator at $H_{bias}$ = 1.12 kOe and (c) shows vortex states for all the shapes at $H_{bias}$ = 0

We have simulated ground states of the samples (s = 100nm) at $H_{bias}$ = 1.12kOe using OOMMF and LLG micromagnetic simulator as shown in fig. 8(a) and (b) respectively. However, $H_{bias}$ = 0, we get vortex state of the samples of s = 100nm (as shown in fig. 8 (c)). We observe that different dot shaped elements forms different magnetization configurations such as onion (O), flower (F), buckle (B) and vortex states with shifted core and different core polarity p = 1 (its core points up) or p = −1 (core points down). We observe that in hexagonal dot array, all the individual nanodots show vortex states while, in diamond and triangular dot array, all individual nanodots do not show vortex states. This is due to different shapes of the dots and particularly the sharpness of the corner of the dot shape which affects the curling of magnetization. As different shapes experiences different magnetostatic field profile (figure 7) and thus give rise to different configuration states.

Table: Effect of different dot shapes on magnetostatic field strength at two different regions of the samples.

| Dot shapes | Magnetostatic field (Oe) | |
|---|---|---|
| | Region A | Region B |
| Triangle | 9151.3 | 409.9 |
| Diamond | 8483.7 | 461.5 |
| Hexagon | 9546.4 | 721.8 |

**Conclusion**

Using micromagnetic simulation tunable spin-wave properties in Py nanodots of different shape for varying inter-dot separation have been investigated. Specifically, triangular, diamond and hexagon dots are studied with varying *s* from 25nm to 500nm. Amongst all shapes, triangular dots show maximum frequency up to 14.7GHz. Variation in *s* up-to 100nm has profound effect on the spin wave spectra. However, beyond 100nm, there is not much variation in SW frequencies. We observed splitting in the modes from single dot to dot array samples. The power profile confirms the nature of the observed SW modes which show quantized mode, centre mode and edge mode like nature. The microscopic origin of the excitation of modes is explained from the contour maps of the dots by the combination of internal magnetic field profiles within the nanodots and the magnetostatic fields within the array. Vortex states are observed in an array for all the shapes at $H_{bias} = 0$ while at $H_{bias} = 1.12$ kOe we observe onion, flower and buckle states in different shaped dot elements. The profound effect of corners in triangle, diamond and hexagon shape magnetic nanodots observe in this work can be useful in achieving tunable control of SW frequencies thereby making it useful in various magnonic device applications.

**Acknowledgement**

The authors acknowledge SGDRI (UPM) project of IIT Kharagpur. Acknowledgement also goes to Dr. A Barman for useful discussion.